\documentclass{ws-procs9x6}

\begin{document}

\title{Measurement of $\sin^2{2\theta_{13}}$
by reactor experiments and its
sensitivity\footnote{\uppercase{T}his work is supported in part by
the \uppercase{G}rant-in-\uppercase{A}id for \uppercase{S}cientific
\uppercase{R}esearch
in \uppercase{P}riority \uppercase{A}reas \uppercase{N}o. 12047222
and \uppercase{N}o. 13640295,
\uppercase{J}apan \uppercase{M}inistry
of \uppercase{E}ducation, \uppercase{C}ulture, \uppercase{S}ports,
\uppercase{S}cience, and \uppercase{T}echnology.}}

\author{Osamu Yasuda}

\address{Department of Physics, Tokyo Metropolitan University,\\
1-1 Minami-Osawa,
Hachioji, Tokyo 192-0397, Japan\\
E-mail: yasuda@phys.metro-u.ac.jp}


\maketitle

\abstracts{
Reactor experiments offer a promising way to
determine $\theta_{13}$ and are free from
parameter degeneracies in neutrino oscillations.
It is described how
reactor measurements of $\sin^22\theta_{13}$ 
can be improved
by a near-far detector complex.
The experimental lower bound is derived on the sensitivity to
$\sin^2{2\theta_{13}}\gtrsim 0.02$ based on the rate analysis,
and an idea is given which may enable us to circumvent this bound.
It is shown that in the Kashiwazaki-Kariwa plan (KASKA)
the sensitivity to $\sin^2{2\theta_{13}}$ is approximately 0.02.
}

\section{Introduction}
The recent experiments on atmospheric and
solar neutrinos and KamLAND have been so successful that
we now know the approximate values of the mixing angles
and the mass squared differences
of the atmospheric and solar neutrino oscillations:
$(\sin^22\theta_{12}, \Delta m^2_{21}\equiv m^2_2-m^2_1)\simeq
(0.8, 7\times10^{-5}{\rm eV}^2)$ for the solar neutrino
and $(\sin^22\theta_{23}, |\Delta m^2_{31}|\equiv |m^2_1-m^2_3|)\simeq
(1.0, 2\times10^{-3}{\rm eV}^2)$ for the atmospheric neutrino,
where the three flavor framework of neutrino oscillations
is assumed.
The quantities which are still unknown are
the third mixing angle $\theta_{13}$, the sign
of the mass squared difference $\Delta m^2_{31}$
which indicates whether the mass pattern of neutrinos is
of normal hierarchy or of inverted one, and the CP phase $\delta$.
Among these, $\theta_{13}$ is the most important quantity
in the near future neutrino experiments.

It has been known
that the oscillation parameters
$\theta_{jk}$, $\Delta m^2_{jk}$, $\delta$ cannot be determined uniquely
even if the appearance probabilities
$P(\nu_\mu\rightarrow\nu_e)$ and
$P(\bar{\nu}_\mu\rightarrow\bar{\nu}_e)$
are measured precisely from a long
baseline accelerator experiment due to so-called parameter 
degeneracies,
and this problem has to be solved
to determine the CP phase in the future long
baseline experiments.
Among the ideas which have been proposed to solve the problem,
combination of a reactor measurement and a long
baseline experiment offers a promising 
possibility (See Ref.\cite{Minakata:2002jv} and references therein).
This combination works because reactor experiments measure the disappearance
oscillation probability
which depends only on $\theta_{13}$ to a good approximation:
\begin{eqnarray}
P(\bar{\nu}_{e} \rightarrow \bar{\nu}_{e}) \simeq 1-
\sin^22\theta_{13}
\sin^2\left(\frac{\Delta m^2_{13}L}{4E}\right),
\nonumber
\end{eqnarray}
where $E$ and $L$ stand for the neutrino energy and the baseline
length.

\section{Reactor measurements of $\sin^2{2\theta_{13}}$
and its sensitivity}
To illustrate how a near-far detector complex
improves the sensitivity to $\sin^22\theta_{13}$
of reactor neutrino experiments,
let me discuss the case with a single
reactor, one near and one far detectors.
To discuss the sensitivity, let me introduce
$\chi^2$, which basically indicates whether
the difference between the number of events with
oscillations and that without oscillations is
large enough compared to the total error
whose square is the sum of the statistical and
systematic errors squared.  For simplicity
I will mainly discuss the sensitivity
in the limit of infinite statistics, i.e.,
with the systematic errors only.
Throughout my talk I will perform the rate analysis only.

Let $m_{\text{n}}$ and $m_{\text{f}}$ be the number of events
measured at the near and far detectors, $t_{\text{n}}$ and $t_{\text{f}}$
be the theoretical predictions.
Then $\chi^2$ is given by
\begin{eqnarray}
\hspace*{-10mm}
\displaystyle
\chi^2&=&\min_{\alpha's}\left\{
\left[{m_{\text{n}}-t_{\text{n}}(1+\alpha_{\text{c}}
+\alpha_{\text{c}}^{\text{(r)}}
+\alpha_{\text{u}}^{(r)})
\over t_{\text{n}}\sigma_{\text{u}}}\right]^2
+\left[{m_{\text{f}}-t_{\text{f}}(1+\alpha_{\text{c}}
+\alpha_{\text{c}}^{\text{(r)}}
+\alpha_{\text{u}}^{(r)})
\over t_{\text{f}}\sigma_{\text{u}}}\right]^2
\right.\nonumber\\
&{\ }&\left.
+\left({\alpha_{\text{c}} \over \sigma_{\text{c}}}\right)^2
+\left({\alpha_{\text{c}}^{\text{(r)}} \over 
\sigma_{\text{c}}^{\text{(r)}}}\right)^2
+\left({\alpha_{\text{c}}^{\text{(r)}} \over 
\sigma_{\text{c}}^{\text{(r)}}}\right)^2\right\},
\label{chi1}
\end{eqnarray}
where $\alpha_u$, $\alpha_c$, $\alpha^{(r)}_c$ and
$\sigma^{(r)}_u$ are the variables to introduce
the uncorrelated systematic error $\sigma_u$ of the
detectors, the correlated systematic error $\sigma_c$
of the detector, the correlated systematic error $\sigma^{(r)}_c$
of the flux and the uncorrelated systematic error $\sigma^{(r)}_u$
of the flux, and
it is assumed that the uncorrelated errors
for the two detectors are the same and are equal to
$\sigma_{\text{u}}$.
After some calculations\cite{multir}, I obtain
\begin{eqnarray}
\hspace*{-20mm}
\displaystyle
\chi^2=
\left(\begin{array}{cc}
\displaystyle{m_{\text{n}} \over t_{\text{n}}}-1, &
\displaystyle{m_{\text{f}} \over t_{\text{f}}}-1
\end{array}
\right)
V^{-1}
\left(\begin{array}{c}
\displaystyle{m_{\text{n}} \over t_{\text{n}}}-1 \\
\\
\displaystyle{m_{\text{f}} \over t_{\text{f}}}-1
\end{array}
\right),
\nonumber
\end{eqnarray}
where
\begin{eqnarray}
V\equiv
\left(
\displaystyle
\begin{array}{ll}
\sigma^2_{\text{u}}+\sigma^2_{\text{c}}
+(\sigma^{\text{(r)}}_{\text{u}})^2
+(\sigma^{\text{(r)}}_{\text{c}})^2
& \quad\sigma^2_{\text{c}}
+(\sigma^{\text{(r)}}_{\text{u}})^2
+(\sigma^{\text{(r)}}_{\text{c}})^2\\
\quad\sigma^2_{\text{c}}
+(\sigma^{\text{(r)}}_{\text{u}})^2
+(\sigma^{\text{(r)}}_{\text{c}})^2 &
\sigma^2_{\text{u}}+\sigma^2_{\text{c}}
+(\sigma^{\text{(r)}}_{\text{u}})^2
+(\sigma^{\text{(r)}}_{\text{c}})^2
\end{array}
\right)\nonumber
\end{eqnarray}
is the covariance matrix.
After diagonalizing $V$, I have
\begin{eqnarray}
\hspace*{-5mm}
\displaystyle
\chi^2=
{\left[\left(m_{\text{n}}/t_{\text{n}}-1\right)
+\left(m_{\text{f}}/t_{\text{f}}-1\right)
\right]^2 
\over 4\sigma^2_{\text{c}}
+4(\sigma^{\text{(r)}}_{\text{u}})^2
+4(\sigma^{\text{(r)}}_{\text{c}})^2
+2\sigma^2_{\text{u}}}
+{\left[\left(m_{\text{n}}/t_{\text{n}}-1\right)
-\left(m_{\text{f}}/t_{\text{f}}-1\right)
\right]^2 
\over 2\sigma^2_{\text{u}}}.
\label{chi2}
\end{eqnarray}
The strategy in this talk is to assume no neutrino oscillation
for the theoretical predictions $t_j$~($j$=n,f) and assume
the number of events with oscillations for the measured values
$m_j$~($j$=n,f) and to examine if a hypothesis with no
oscillation is excluded, say at the 90\%CL, from the value
of $\chi^2$.
Hence I have
\begin{eqnarray}
\displaystyle
{m_j \over t_j}-1 = -\sin^22\theta_{13}
\left\langle \sin^2\left({\Delta m^2_{13}L_j \over 4E}
\right) \right\rangle,
\label{cond0}
\end{eqnarray}
where $L_j$ is the distance between the reactor and the near or far
($j$=n,f) detector, and
\begin{eqnarray}
\displaystyle
\left\langle \sin^2\left({\Delta m^2_{13}L_j \over 4E}
\right) \right\rangle\equiv
{\displaystyle
\int dE~\epsilon(E)f(E)\sigma(E)\sin^2\left({\Delta m^2_{13}L_j \over 4E}
\right) \over \displaystyle
\int dE~\epsilon(E)f(E)\sigma(E)}.
\nonumber
\end{eqnarray}
$\epsilon(E)$, $f(E)$, $\sigma(E)$ stand for the detection
efficiency, the neutrino flux, and the cross section, respectively.
\begin{figure}
\vglue -0.5cm
\hglue 1.7cm
\includegraphics[scale=0.45]{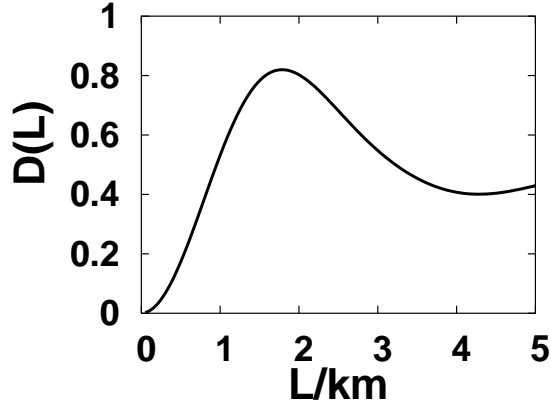}
\vglue -1.0cm
\caption{
$D(L)\equiv \langle
\sin^2\left({\Delta m^2_{13}L \over 4E}\right)\rangle$
as a function of $L$.  $D(L)$ has its maximum value 0.82 at
$L$=1.8km, and approaches to its asymptotic value 0.5
as $L\rightarrow\infty$.
$|\Delta m^2_{13}|=2.5\times10^{-3}$eV$^2$
is used as the reference value.
}
\label{fig0}
\end{figure}
Thus Eq. (\ref{chi2}) becomes
\begin{eqnarray}
\chi^2=\sin^42\theta_{13}\left\{
{\left[D(L_{\text{f}})+D(L_{\text{n}})\right]^2 \over
4\sigma^2_{\text{c}}+4(\sigma^{\text{(r)}}_{\text{u}})^2
+4(\sigma^{\text{(r)}}_{\text{c}})^2
+2\sigma^2_{\text{u}}}
+{\left[D(L_{\text{f}})-D(L_{\text{n}})\right]^2 \over 
2\sigma^2_{\text{u}}}\right\},
\label{chi3}
\end{eqnarray}
where (\ref{cond0}) was used, and
\begin{eqnarray}
\displaystyle
D(L)\equiv\left\langle \sin^2\left({\Delta m^2_{13}L \over 4E}
\right) \right\rangle\nonumber
\end{eqnarray}
was defined.  The numerical value of $D(L)$ is plotted in Fig.\ref{fig0}
as a function of $L$.
Here I adopt the reference values 
$\sigma_{\text{c}}=0.8\%/\sqrt{2}=0.6\%$ and
$\sigma_{\text{u}}=\sqrt{(2.7\%)^2-(2.1\%)^2-(0.8\%/\sqrt{2})^2}=1.6\%$
used in Ref.\cite{Minakata:2002jv}, where basically the
reference values were deduced from extrapolation of the previous
reactor experiments Bugey and CHOOZ.
In the estimation of $\sigma_{\text{c}}$,
I used 2.7\% total error and 2.1\% error of the flux
which are the reference values in the CHOOZ experiment.
As for the correlated and uncorrelated errors
of the the flux from the reactors,
I adopt the same reference values as those used by
the KamLAND experiment:
$\sigma_{\text{c}}^{\text{(r)}}=2.5\%$,
$\sigma_{\text{u}}^{\text{(r)}}=2.3\%$.
Putting the present reference values together, I have
\begin{eqnarray}
\hspace*{-10mm}
2\sigma_{\text{u}}^2&=&(0.8\%)^2\nonumber\\
4\sigma_{\text{c}}^2+4(\sigma^{\text{(r)}}_{\text{u}})^2
+4(\sigma^{\text{(r)}}_{\text{c}})^2+2\sigma_{\text{u}}^2&=&
(7.6\%)^2.\nonumber
\nonumber
\end{eqnarray}
The contribution from $(4\sigma_{\text{c}}^2
+4(\sigma^{\text{(r)}}_{\text{u}})^2
+4(\sigma^{\text{(r)}}_{\text{c}})^2
+2\sigma_{\text{u}}^2)^{-1}$
in Eq. (\ref{chi3}) is only 1\% compared to that from
$(2\sigma_{\text{u}}^2)^{-1}$, so virtually this term can be ignored
in Eq. (\ref{chi3}).
Hence $\chi^2$ is given approximately by
\begin{eqnarray}
\chi^2\simeq\sin^42\theta_{13}
{\left[D(L_{\text{f}})-D(L_{\text{n}})\right]^2 \over 
2\sigma^2_{\text{u}}}.
\label{chi4}
\end{eqnarray}
Since no oscillation is assumed for the theoretical
predictions, the best fit value in $\sin^22\theta_{13}$
for the measurement is $\sin^22\theta_{13}=0$.
If $\chi^2$ is larger than 2.7, which
corresponds to the value at the 90\%CL for one degree
of freedom, then the hypothesis of no oscillation is
excluded at the 90\%CL.  This implies that the systematic limit
on $\sin^22\theta_{13}$ at the 90\%CL, or the sensitivity
in the limit of infinite statistics, is given by
\begin{eqnarray}
\left(\sin^22\theta_{13}\right)_{\text{limit}}^{\text{sys~only}}
\simeq\sqrt{2.7}{\sqrt{2}\sigma_{\text{u}} \over 
D(L_{\text{f}})-D(L_{\text{n}})}.
\label{sens0}
\end{eqnarray}
Eq. (\ref{sens0}) tells us that, in order to optimize
$\left(\sin^22\theta_{13}\right)_{\text{limit}}^{\text{sys~only}}$,
I have to minimize $D(L_{\text{n}})\equiv
\langle \sin^2\left({\Delta m^2_{13}L_{\text{n}} / 4E}
\right)\rangle$ and maximize $D(L_{\text{f}})\equiv
\langle \sin^2\left({\Delta m^2_{13}L_{\text{f}} / 4E}
\right)\rangle$.  Since the possible maximum value of
$D(L_{\text{f}})-D(L_{\text{n}})$ is 0.82,
which is attained for $L_{\text{f}}=1.8$km and $L_{\text{f}}=0$,
the lower bound of
$\sin^22\theta_{13}$
at a single reactor experiment can be estimated as:
\begin{eqnarray}
\mbox{\rm lower bound of }
\sin^22\theta_{13}
\simeq{\sqrt{2.7}\sqrt{2} \sigma_{\text{u}}\over 0.82}
=2.8\,\sigma_{\text{u}}.
\label{lbound1}
\end{eqnarray}
Eq. (\ref{lbound1}) indicates that,
unless one develops a technology to improve $\sigma_{\text{u}}$
significantly compared to the reference value
$\sigma_{\text{u}}=0.6\%$ assumed in Ref.\cite{Minakata:2002jv},
it is difficult to achieve the sensitivity below
$\sin^22\theta_{13}=0.016$ in an experiment with one reactor and
two detectors.
It should be stressed here that the correlated systematic error
$\sigma_{\text{c}}$ and the errors
$\sigma^{\text{(r)}}_{\text{u}}$,
$\sigma^{\text{(r)}}_{\text{c}}$ of the flux
do not appear in the dominant contribution
to $\chi^2$, and that it is the uncorrelated systematic
error $\sigma_{\text{u}}$ divided by the
factor $D(L_{\text{f}})-D(L_{\text{n}})$
that determines the systematic limit
on $\sin^22\theta_{13}$.

In the ideal case with $N$ pairs of the reactors and the near
detectors and one far detector, one can show\cite{multir}
that Eq. (\ref{lbound1})
becomes
\begin{eqnarray}
\mbox{\rm lower bound of }
\sin^22\theta_{13}
\simeq{\sqrt{2.7}\sqrt{1+1/N} \sigma_{\text{u}}\over 0.82}
> 2.0\,\sigma_{\text{u}}.
\label{lbound2}
\end{eqnarray}
Eq.\. (\ref{lbound2}) indicates that even in the ideal case
the sensitivity can be approximately 0.012 at best,
as far as the rate analysis is concerned,
no matter how many reactors and detectors there may be,
if one adopts $\sigma_{\text{u}}=0.6\%$.

In principle there is a way to improve this bound.
In the case with one reactor,
if one puts $M$ identical detectors at the near site
and $M$ identical detectors at the far site, where all these detectors
are assumed to have the same uncorrelated systematic error
$\sigma_{\text{u}}$, then the value of $\chi^2$ is simply
multiplied by $M$.  Eqs. (\ref{chi4}) and (\ref{lbound1})
imply that the sensitivity becomes
\begin{eqnarray}
\mbox{\rm lower bound of }
\sin^22\theta_{13}
\simeq{\sqrt{2.7}\sqrt{2} \sigma_{\text{u}}\over 0.82 \sqrt{M}}
={2.8 \over \sqrt{M}} \,\sigma_{\text{u}}.
\nonumber
\end{eqnarray}
Therefore, it follows theoretically that the more identical detectors
one puts, the better sensitivity one gets.  Notice that this conclusion
is based crucially on the assumption that the uncorrelated systematic error
$\sigma_{\text{u}}$ of the detectors is independent of the number of
the detectors $2M$.
This assumption may not be satisfied in general,
but if the dependence of $\sigma_{\text{u}}$ on $M$ is
weaker than $\sqrt{M}$, $\sigma_{\text{u}}/\sqrt{M}$ decreases
as $M$ increases, and this possibility may give us a way to improve
the sensitivity.  It is therefore important to examine experimentally
the dependence of $\sigma_{\text{u}}$ on $M$.

\begin{figure}
\vglue -1.0cm
\hglue -0.3cm
\includegraphics[scale=0.25]{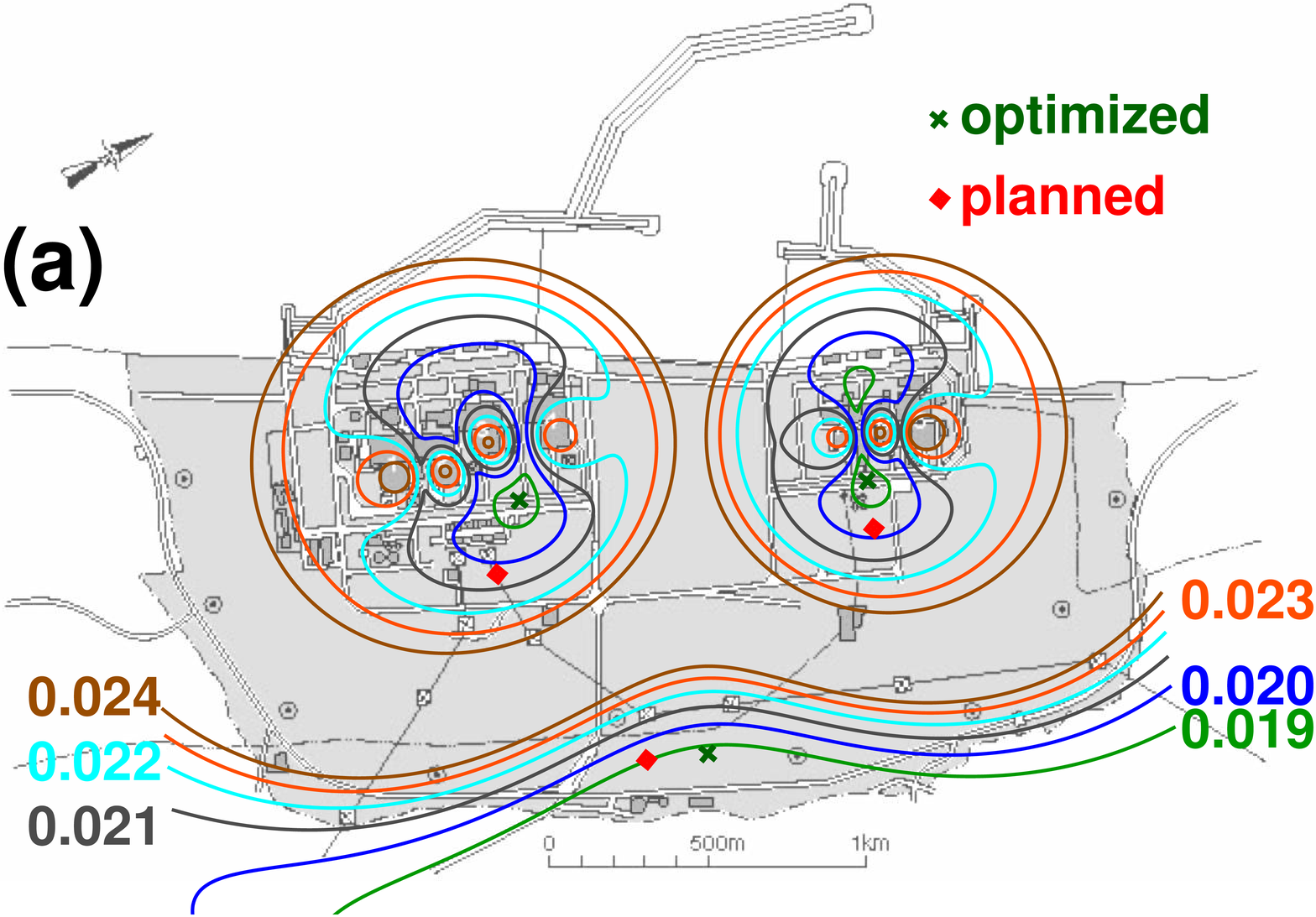}
\hglue -0.5cm
\includegraphics[scale=0.25]{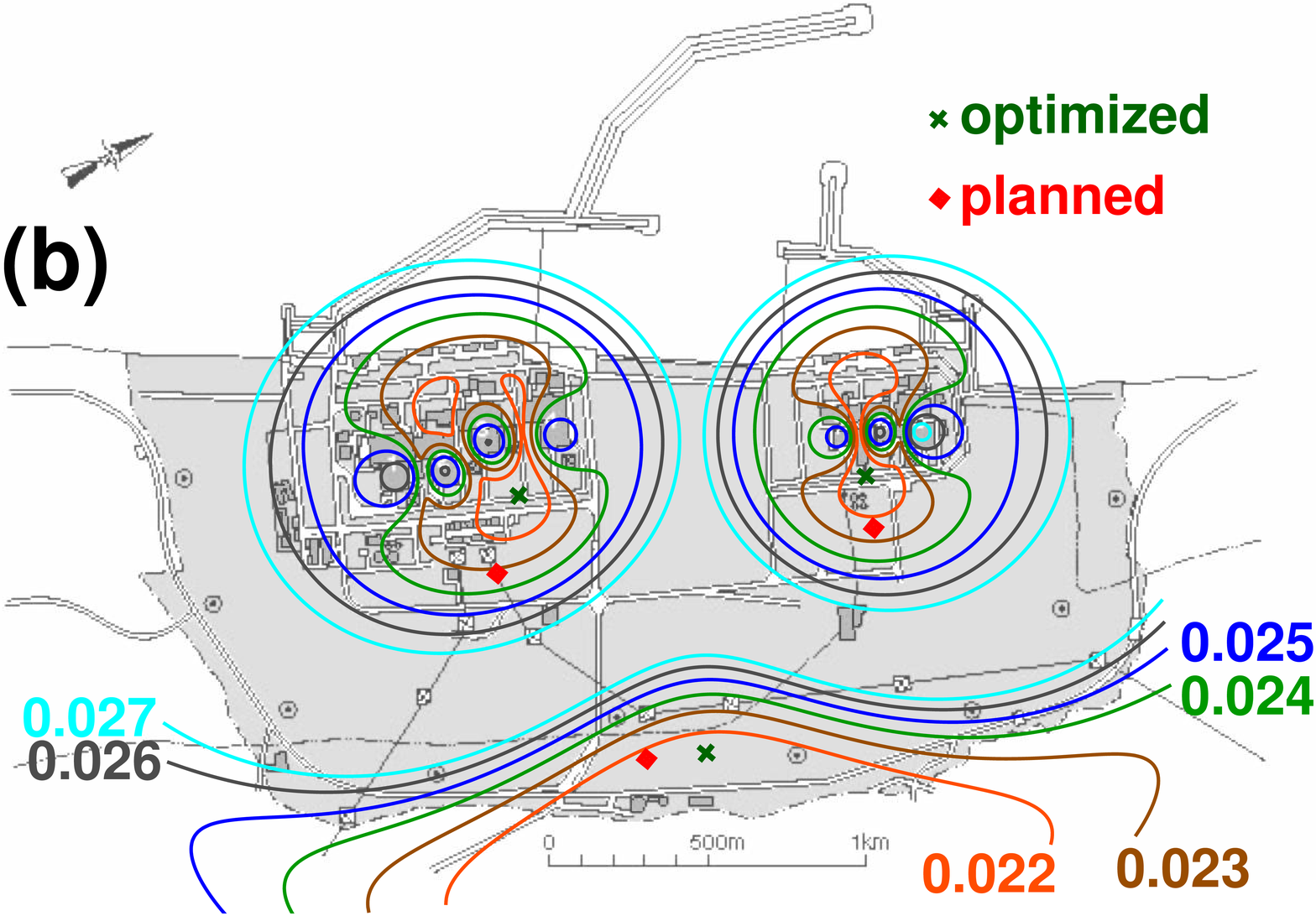}
\caption{
The contour plot of the sensitivity
to $\sin^22\theta_{13}$ in the limit of
infinite statistics (a) and for 20 ton$\cdot$yr (b) in the KASKA experiment.
The optimized and currently planned positions of the detectors
are also depicted.
When the contour for each detector is plotted,
it is assumed that other detectors are located in
the optimized positions.  The position of the far detector is
varied within the campus of the Kashiwazaki-Kariwa nuclear plant.
For the data size 20 ton$\cdot$yr,
the value $\sin^22\theta_{13}= 0.025$ of the sensitivity
seems to be inconsistent with
the contour plot in Fig.\ref{figkk}(b), which suggests that it
is better than $\sin^22\theta_{13}= 0.022$ for the
planned location of the far detector.  However,
the contour for the far detector
is plotted here on the assumption that the near detectors are both
in the optimized position, and therefore this calculation
gives the sensitivity better than that for the
locations currently planned.}
\label{figkk}
\end{figure}

\section{The KASKA plan\label{kaska}}
The Kashiwazaki-Kariwa nuclear plant consists of two clusters
of reactors, and one cluster consists of four reactors while the
other consists of three.
In the Kashiwazaki-Kariwa (KASKA) plan\cite{Suekane:2003nh}
one near detector is placed near one cluster of the reactors
while the other near detector is placed near another cluster.
In addition, a far detector is located \mbox{1.3\,km} away from the reactors.
In the presence of seven reactors, the total number $m_j$ of the
events measured at the $j$-th detector ($j$=1 (the first near),
2 (the second near), 3 (the far detector))
is a sum of contributions $m_{aj}~(a=1,\cdots,7)$ from
each reactor, and this is also the case for the
theoretical predictions $t_j$
and $t_{aj}~(a=1,\cdots,7)$ at the $j$-th detector.
So I have
$m_j=\sum_{a=1}^7 m_{aj}$,
$t_j=\sum_{a=1}^7 t_{aj}$.
In the limit of infinite statistics $\chi^2$ is defined as
\begin{eqnarray}
\hspace*{-20mm}
\chi^2&=&\min_{\alpha's}\left\{
\displaystyle\sum_{j=1}^3
{1 \over t_j^2\sigma_{\text{u}}^2}
\left[m_j-t_j\left(1+\alpha_{\text{c}}+\alpha_{\text{c}}^{\text{(r)}}
+\sum_{a=1}^7
{t_{aj} \over t_j} \alpha_{\text{u}a}^{(r)}\right)\right]^2\right.
\nonumber\\
&{\ }&+\left.\left({\alpha_{\text{c}} \over \sigma_{\text{c}}}\right)^2
+\left({\alpha_{\text{c}}^{\text{(r)}} \over 
\sigma_{\text{c}}^{\text{(r)}}}\right)^2
+\sum_{a=1}^7 \left({\alpha_{\text{u}a}^{\text{(r)}} \over 
\sigma_{\text{u}}^{\text{(r)}}}\right)^2\right\}.
\nonumber
\end{eqnarray}
After some calculations, I get
\begin{eqnarray}
\chi^2
&=&\left(\displaystyle{m_1 \over t_1}-1,\displaystyle{m_2 \over t_2}-1,
\displaystyle{m_3 \over t_3}-1\right)
V^{-1}
\left(
\displaystyle{m_1 \over t_1}-1,\displaystyle{m_2 \over t_2}-1,
\displaystyle{m_3 \over t_3}-1
\right)^T,
\nonumber
\end{eqnarray}
where
\begin{eqnarray}
V_{jk}&=&\delta_{jk}\sigma_{\text{u}}^2
+\sigma_{\text{c}}^2
+(\sigma_{\text{c}}^{\text{(r)}})^2
+(\sigma_{\text{u}}^{\text{(r)}})^2\displaystyle 
\sum_{a=1}^7 \displaystyle{t_{aj} \over t_j}
{t_{ak} \over t_k}.
\label{v}
\end{eqnarray}
As in the case with one reactor (\ref{cond0}), I have
\begin{eqnarray}
\displaystyle
{m_j \over t_j}-1 = -\sin^22\theta_{13}
\sum_{a=1}^7 \displaystyle{t_{aj} \over t_j}
D(L_{aj}),
\nonumber
\end{eqnarray}
where $L_{aj}$ is the distance between the $a$-th reactor
($a$=1,$\cdots$,7) and the $j$-th detector ($j$=1,2,3).
Therefore $\chi^2$ is proportional to $\sin^42\theta_{13}$:
\begin{eqnarray}
\chi^2=C\sin^42\theta_{13},
\nonumber
\end{eqnarray}
where
\begin{eqnarray}
C\equiv\sum_{j,k=1}^3
\sum_{a=1}^7 \displaystyle{t_{aj} \over t_j}D(L_{aj})
(V^{-1})_{jk}
\sum_{b=1}^7 \displaystyle{t_{bk} \over t_k}D(L_{bk}).
\nonumber
\end{eqnarray}
Hence the sensitivity to $\sin^22\theta_{13}$ at 90\%CL is
given by
\begin{eqnarray}
\sin^22\theta_{13}&=&\left({\left.\chi^2\right|_{90\%CL} \over C}\right)^{1/2}
=\left({2.7 \over C}\right)^{1/2},
\nonumber
\end{eqnarray}
where it is assumed that the value of $|\Delta m^2_{13}|$ is precisely
known and therefore degree of freedom in this analysis is one.
The covariance matrix (\ref{v}) cannot be inverted
analytically, so $C$ has to be evaluated numerically.
Fig.\ref{figkk}
shows the contour plot of the sensitivity
to $\sin^22\theta_{13}$ in the limit of
infinite statistics (a) and for 20 ton$\cdot$yr (b).
Fig.\ref{figkk} indicates that optimization forces
the distances
between each near detector and the reactors in
each cluster be approximately (300$\pm$130)m.  This results
in slightly poorer sensitivity to $\sin^22\theta_{13}$
than the hypothetical single reactor case with
near and far detectors.  This is because in the case with
a single reactor, the near detector can be
theoretically arbitrarily close to the reactor
and the factor
$D(L_{\text{f}})-D(L_{\text{n}})\rightarrow D(L_{\text{f}})$
can in principle be the maximum value 0.82, whereas
in the KASKA case, the sensitivity is given by\cite{multir}
\begin{eqnarray}
\hspace*{-10mm}
&{\ }&\left(\sin^22\theta_{13}\right)_{\text{limit}}^{\text{sys~only}}
\nonumber\\
&\simeq&{\sqrt{2.7}\sqrt{1.04}~\sigma_{\text{u}} \over 
0.81\displaystyle\sum_{a=1}^7 \displaystyle{t_{a3} \over t_j}D(L_{a3})
-0.31\sum_{a=1}^7 \displaystyle{t_{a1} \over t_j}D(L_{a1})
-0.50\sum_{a=1}^7 \displaystyle{t_{a2} \over t_j}D(L_{a2})},
\nonumber
\end{eqnarray}
and the second and third term in the denominator
give small contribution to spoil the sensitivity.
From the numerical calculations, one can show that
the KASKA experiment has the sensitivity
$\sin^22\theta_{13}\sim 0.025$ (0.019) with the data size of 20 ton$\cdot$yr
($\infty$ ton$\cdot$yr).
Since the sensitivity to $\sin^22\theta_{13}$ at KASKA is
close to the lower bound 0.016
(cf. (\ref{lbound1}) with $\sigma_{\text{u}}=0.6\%$),
the setup of the KASKA plan is not far from the optimum.

\end{document}